# Monolithic Expandable-FOV Metalens Enabled by Radially Gradient-Tilted Meta-Atoms

Feiyang Zhang[1], Guoxia Han[1,*], Yihan Tian[1], Yanbin Ma[1], Xianghua Yu[2], Xiaolong Liu[3]

1, College of Science, China University of Petroleum (East China), Qingdao 266580, Shandong, China

2, State Key Laboratory of Ultrafast Optical Science and Technology, Xi'an Institute of Optics and Precision Mechanics, Chinese Academy of Sciences, Xi'an, Shaanxi, China

3, Department of Electronics and Nanoengineering, Aalto University, 02150 Espoo, Finland

**\*Corresponding author**

E-mail address：gxhan@upc.edu.cn    **(G. Han)**

**Abstract:** Metalens, as the most promising and applicable emerging optical device, has long been constrained by the limited field of view (FOV). Recent studies employing phase engineering or multi-layer strategies have made some progress, but they all rely on upright meta-atoms. This leads us to consider whether tilted meta-atoms could represent a promising yet underexplored approach for enhancing the field of view? In this work, we introduce the control of the tilt angle of meta-atoms as a new degree of freedom into the design of metalenses and propose a wide-field-of-view (WFOV) metalens design framework utilizing radially gradient-tilted meta-atoms. Based on the proposed method, we designed two WFOV metalenses with distinct tilt configurations to meet specific application requirements of efficiency or precision. Simulation results demonstrate that both designs, exhibiting distinct performance characteristics, achieve diffraction-limited focusing within a 120° FOV (±60°). Additionally, the FOV can be expanded with this method by tuning the tilt angle configurations of meta-atoms and the metalens diameter. These findings will establish a promising pathway towards compact optical systems capable of combining ultra-wide angular coverage with high-resolution imaging.

**Keywords:** metalens; wide field of view; tilted meta-atoms; imaging.

# 1. Introduction

Metalenses [1-3], a revolutionary class of optical components, utilize sub-wavelength meta-atom arrays with precisely engineered phase profiles to enable multi-dimensional control of light fields. This includes amplitude modulation [4, 5], phase engineering [6, 7], and polarization manipulation [8, 9]. Compared to conventional refractive optics, metalenses offer superior performance characteristics, such as sub-millimeter thickness, reduced weight, and discrete phase control at each meta-atom, enabling adaptive control over spatiotemporal light fields across multiple wavelengths [10-12]. These attributes make metalenses ideal for integration into compact optical systems, facilitating applications in military surveillance, industrial automation, biomedical diagnostics, and consumer-grade imaging technologies [13-17].

Recent advancements in metalens functionalities, including zoom imaging [18, 19], broadband chromatic aberration correction [20-22], and wide-field-of-view (WFOV) imaging, have significantly broadened the application scope. Notably, WFOV imaging in the visible spectrum (VIS) band has gained particular attention for its high-resolution clarity and natural color reproduction [23], driving critical innovations in wearable VR/XR headsets, ultra-high-definition holography, and smart city infrastructure for traffic and environmental analytics.

Despite these advances, achieving high-fidelity imaging over a truly WFOV remains a formidable challenge. Off-axis aberrations, chiefly coma, astigmatism and field curvature, intensify dramatically as the angle of incidence increases, degrading resolution and contrast toward the edge of the field. To address this limitation, three primary strategies have emerged: multilayered metalenses [24, 25], quadratic phase modulation [26, 27] and meta-atom structural redesign [32]. Under the first strategy, Arbabi et al. [28] developed a WFOV metalens using a double-layer metasurface architecture, where the first layer mimics the functionality of a Schmidt corrector plate to converge central rays and diverge edge rays, while the second layer focuses the beam. This design achieves a 60° FOV with 70% diffraction efficiency. Similarly, Lin et al. [29] proposed a WFOV metalens through topology-optimized multi-layer aperiodic structures, using abundant optimization parameters to improve modulation of large-angle light. However, multi-layer configurations, while improving phase control flexibility and easing design constraints on individual layers, increase structural complexity and compromise the flatness of metalenses.

Another approach involves quadratic phase modulation, as first proposed by Pu et al. in 2017 [30]. This approach compensates for the linear phase gradient induced by oblique incidence through spatially varying quadratic phase modulation, converting the obliquity of incident light into a lateral focal spot shift. This effectively decouples focusing efficiency from the incident angle and suppresses off-axis aberrations. However, this compensation is strictly valid only under paraxial conditions, and in the non-paraxial regime, the light fails to meet the quadratic phase focusing criterion, resulting in defocused background noise that degrades image contrast. To mitigate the noise, Yang F et al. [31] introduced a spatial filtering aperture to block the non-paraxial rays, reducing noise at the expense of optical throughput. Nonetheless, quadratic phase profiles inevitably introduce spherical aberrations and accurate imaging is achievable only when the focal length significantly exceeds the metalens dimensions to minimize spherical aberrations. Furthermore, due to the upright structure, these approaches suffer from a progressive degradation of meta-atom coupling efficiency under large angle incidence, fundamentally limiting their performance in WFOV scenarios.

Interestingly, Wang et al. [32] introduced a uniform tilt angle to the meta-atoms in the sub-metalens for a specific incident angle and attained a high-efficiency focusing. However, this approach requires incorporating multiple sub-metalenses for different incident angle and stitching their sub-FOVs to realize the WFOV. Therefore, this methodology will introduce sub-image registration errors and hinder the fabrication of compact device inevitably. These limitations on existing strategies present a critical gap: the need for a simple, effective approach to manage large-angle light without compromising device elegance or performance.

To overcome the inherent limitations of uniformly tilted meta-atom arrays, we introduce a design framework based on globally programmed tilt-angle gradients across a single, monolithic metasurface. In contrast to prior approaches that rely on discrete sub-arrays or stitched elements, our method independently tunes the tilt angle of each individual meta-atom to match its local angle of incidence, thereby enabling efficient and aberration-minimized focusing across the full aperture. As shown in Fig. 1(a), the radially varying tilt angles provide localized phase compensation tailored to the angle of incidence, while a co-designed aperture selectively blocks aberration-prone rays, ensuring that incident light from a broad range of angles is efficiently directed to the intended focal region. We systematically investigate focusing characteristics, focusing efficiency, and imaging fidelity of the designed metalens. Simulations at a wavelength of 633 nm confirm diffraction-limited performance across a 120° FOV. By introducing radially gradient-tilted meta-atoms as a new design paradigm, this work establishes a powerful framework for constructing wide-field, high-resolution metalenses, offering a promising route toward compact, high-performance optical systems.

## 2. Methods and Theory

The key to achieving a WFOV metalens lies in precisely controlling the phase of meta-atoms, the sub-wavelength structures composing the metasurface. Unlike conventional quadratic phase profiles, which introduce spherical aberration under oblique incidence, the hyperbolic phase profile excels in wide-angle operation by inherently suppressing off-axis wavefront distortion through its non-paraxial diffraction formulation. In a canonical metalens design, the hyperbolic phase profile is mathematically expressed as [2, 33]:

$$\phi_0(x,y) = \frac{2\pi}{\lambda}(f - \sqrt{x^2 + y^2 + f^2}), \tag{1}$$

where $x$ and $y$ are the meta-atoms coordinates, $\lambda$ is the wavelength, and $f$ is the focal length of metalens.

While the hyperbolic phase profile offers superior aberration control for collimated beams under paraxial incidence, its conventional formulation is designed for light incidence along the optical axis—a condition rarely met in WFOV operation. For oblique incidence (angle $\theta \neq 0°$), the wavefront tilt introduces an additional phase term proportional to $\sin(\theta)$, altering the required phase compensation. This angle-dependent modification results in the following phase distribution:

$$\phi(x, y) = -\frac{2\pi}{\lambda}[\sqrt{(x-x_0)^2 + (y-y_0)^2 + f^2} - \sqrt{x_0^2 + y_0^2 + f^2} + (x\sin\theta_x + y\sin\theta_y)], \quad (2)$$

where $x_0$ and $y_0$ stand for the focal point coordinates in the image plane, and $\theta_x$ and $\theta_y$ represent the incident angle components along the x- and y-directions, respectively.

Metasurfaces manipulate wavefronts through two primary mechanisms: propagation phase and geometric (Pancharatnam-Berry) phase. The propagation phase arises from optical path differences determined by meta-atom geometry, while the geometric phase results from rotating asymmetric meta-atoms to manipulate polarization conversion. In this work, we utilize the propagation phase for its polarization-insensitive properties, achieved using rotationally symmetric meta-atoms. The phase delay is tuned by adjusting the structural parameters, such as the height and radius of each meta-atom, which govern the effective refractive index.

Specifically, we employ elliptical titanium dioxide (TiO$_2$) nanopillars on a silica (Si) substrate as the fundamental meta-atoms. As illustrated in Fig. 1(b), each nanopillar is defined by its height $H$, radius $d$, and lattice period $P$. The period $P$ is designed to satisfy the Nyquist sampling criterion ($P < \lambda/2NA$, where NA is the Numerical Aperture), ensuring adequate sampling of the phase profile to prevent aliasing artifacts. For optimal performance at a wavelength of 633 nm, chosen for its high resolution in the VIS band, we set the $H = 0.5$ μm and $P = 0.4$ μm, respectively. To enable WFOV imaging, each meta-atom is tilted at an angle $\alpha$ relative to the optical axis, as depicted in Fig. 1(c). Although the overall shape of the meta-atom becomes relatively irregular, such structures can be fabricated using methods such as distributed lithography or nanoimprint lithography in our investigation. Using the Finite-Difference Time-Domain (FDTD) method with periodic boundary conditions, we simulated the optical properties of the meta-atoms under X-polarized plane wave illumination at 633 nm. The selected radius range is from 0.02 micrometers to 0.2 micrometers, with 38 sampling points calculated within this range, which is feasible for current fabrication technologies. The resulting transmittance and phase, plotted against nanopillar radius in Fig. 1(d), demonstrate full 0–2π phase coverage and possess high average transmittance.

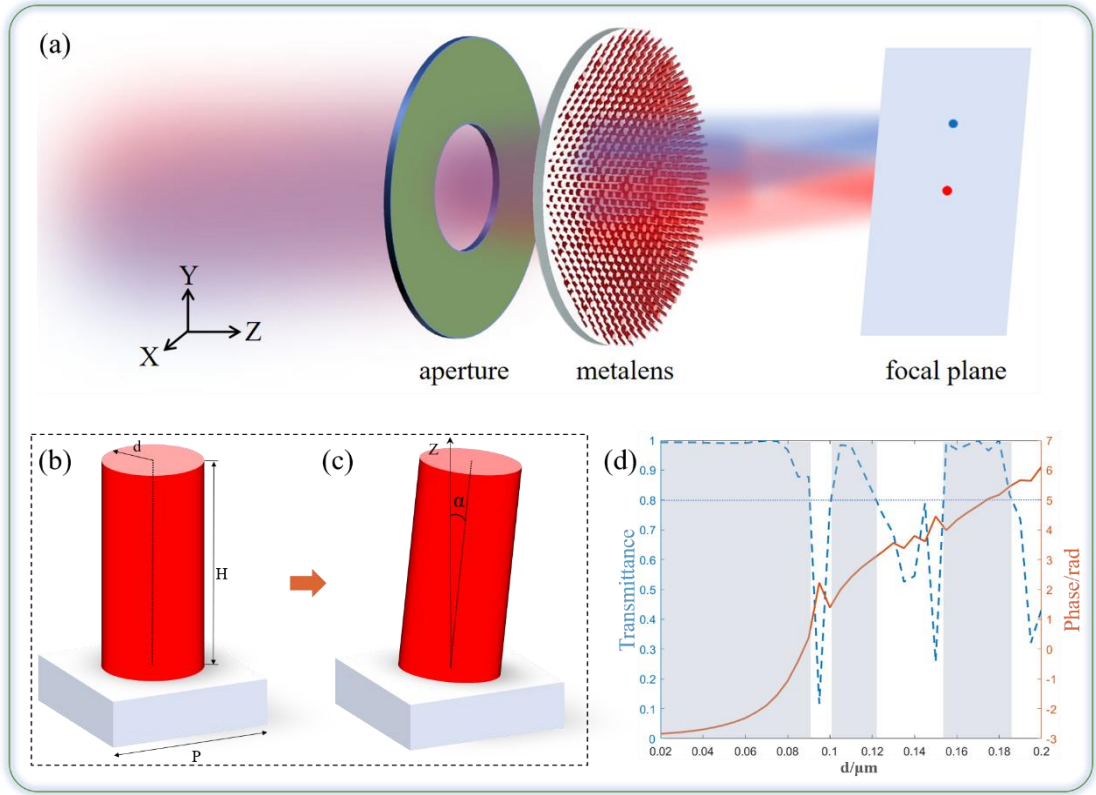

**Fig. 1.** Metalens design and meta-atom library. (a) Schematic of the WFOV metalens imaging system based on radial gradient-tilt of meta-atoms. 3D view of a (b) normal and (c) tilted meta-atom on a silica substrate. (d) Phase (red curve) and transmittance (blue curve) as functions of nanopillar radius at 633 nm.

For each position on the substrate, the required focusing phase was derived from the hyperbolic profile (Eq.2). Meta-atom dimensions were selected using a nearest-neighbor phase-matching algorithm, matching the target phase to the closest value in a precomputed database. These dimensions were encoded onto the substrate, laying the groundwork for the optical performance of the metalens for WFOV imaging.

To address these challenges, we propose a radially tilted meta-atom array that simultaneously corrects off-axis wavefront distortion and enhances angular response. By introducing tilt angles that vary gradually along the radial direction, each meta-atom imparts an additional phase term to counteract wavefront tilting caused by oblique incidence.

## 3. Results and Discussion

Conventional meta-atoms suffer from efficiency degradation and phase errors at large incident angles, greatly limiting WFOV performance. In contrast, meta-atoms with tilt angle configuration can impart additional phase terms to counteract wavefront tilting caused by oblique incidence and these configurations may effectively improve the coupling efficiency of lights. As evidence, the focusing fields of conventional metalens and the metalens whose meta-atoms all possessing tilt angle $\alpha = 20°$ are calculated under the X-polarized plane wave of $\theta = 0°$ and $20°$ and shown in Fig. 2, respectively.

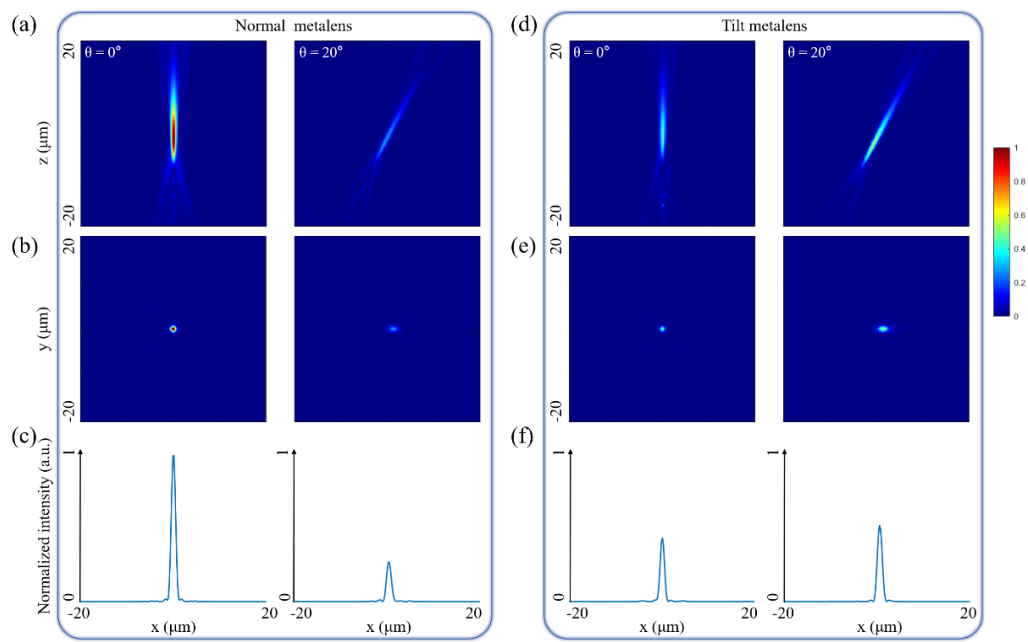

**Fig. 2.** (a) The axial light field, (b) focal plane field, and (c) intensity distribution curve along the x-axis at the focal plane of a conventional metalens and the (d) axial light field, (e) focal plane field, and (f) intensity distribution curve along the x-axis at the focal plane of a metalens with all meta-atoms having a 20° tilt angle under 0° and 20° plane wave illumination, respectively.

It can be seen that the metalens with tilted meta-atom configuration has higher focus energy and fewer secondary diffraction spots than the traditional metalens under 20° of illumination and its focus energy under the tilt angle is greater than that under the vertical illumination of light. This proves the effectiveness of the tilted configurations of meta-atoms. Meta-atoms with tilt angles exhibit more "precise and robust" control when illuminated by light with an incident angle close to their orientation, similar to how conventional metalenses achieve optimal focusing performance under on-axis illumination.

To further tap the potential of tilt configuration and address the trade-off between optical throughput and imaging precision in WFOV systems, by employing radially gradient-tilted meta-atoms, we propose two design paradigms: a Position-Dependent Tilted-Meta-atom (PDTM) metalens optimized for high energy throughput, and an Aperture-Dependent Tilted-Meta-atom (ADTM) metalens for enhanced imaging precision. These designs leverage strategic tilt-angle engineering to achieve wide-angle performance. In this section, all light sources utilized X-polarized plane waves, and the simulations were conducted using the FDTD method with perfectly matched layer (PML) boundary conditions.

## 3.1 Position-Dependent Tilted Meta-atom (PDTM) Metalens

The PDTM employs a position-dependent tilt-angle strategy to maximize energy throughput across a wide angular range, defined by Equation (3):

$$\alpha_1 = \theta_0 \cdot \frac{r}{R}, \tag{3}$$

where $\alpha_1$ is the tilt angle, $\theta_0 = 70°$ is half the maximum FOV angle, $r$ is the radial position of meta-atom, and $R = 16$ μm is the metalens radius. This linear tilt distribution ensures that meta-atoms at larger radii compensate for steeper wavefront tilts under oblique incidence. Fig. 3(a) illustrates the sectional structure of the PDTM, with a NA of 0.625, a diameter of 32 μm and a focal length of 20 μm, respectively. The z-axis aligns with the optical axis, normal to the metalens surface.

To characterize the focusing performance of PDTM, we simulated the light intensity distribution for incident angles from 0° to 60°. Fig. 3(b-c) show the intensity distributions in the focal and axial planes, respectively. The PDTM achieves effective focusing across the 120° FOV, confirming the efficacy of the tilted configuration. Fig. 3(d) compares normalized intensity profiles along the *x*-axis in the focal plane for various angles. At 30°, the peak intensity drops to 37.2% of the 0° value, indicating reduced light concentration at larger angles due to phase mismatches or reduced coupling efficiency. Despite this, the focal plane exhibits negligible secondary diffraction orders or stray light ensuring clean focusing.

Then, we come to considering the imaging performance of PDTM and three important parameters should be simulated and investigated: the total transmittance, diffraction efficiency, and Strehl ratio. The total transmittance is defined as the ratio of the integrated energy at the focal plane to the total energy of the incident light. Diffraction efficiency is the ratio of the integrated energy within a circular region centered on the focal point with a radius equivalent to three times the full width at half maximum (FWHM) to the total integrated energy at the focal plane. The Strehl ratio is defined as the ratio of the idea FWHM of the focal spot to the FWHM of the actual optical system. The results of these three parameters for PDTM are shown in Fig. 3(e). From the figure, we can see that the transmittance (average 0.42) and diffraction efficiency (average 0.39) remain independent of the incident angle, a desirable trait for consistent light throughput and efficient redirection into the primary diffraction order. The Strehl ratio decreases linearly with angle, reaching 0.48 at 30°, suggesting that aberrations (e.g., coma, astigmatism) slightly degrade focus quality at larger angles. Nevertheless, a Strehl ratio above 0.4 supports acceptable imaging performance for many applications. Additionally, the modulation transfer function (MTF) is calculated to evaluate the imaging performance of the metalens, as illustrated in Fig. 3(f). This is achieved by computing the point spread function (PSF) at multiple field angles, followed by a Fourier transform of the PSF results—the modulus of this transformation yields the MTF. The resulting curves demonstrate excellent on-axis (0°) performance, indicating high resolution and contrast at the center of the field. However, off-axis performance degrades significantly with increasing field angle, with MTF values dropping rapidly for angles beyond 30°, particularly at higher spatial frequencies. For a more intuitive comparison, we provide a representative selection of each of the WFOV metalens architectures and summarize the comparison between them in Table 1. It can be observed that the PDTM can achieve an excellent level of both FOV and efficiency, while also exhibiting a relatively larger numerical aperture (NA).

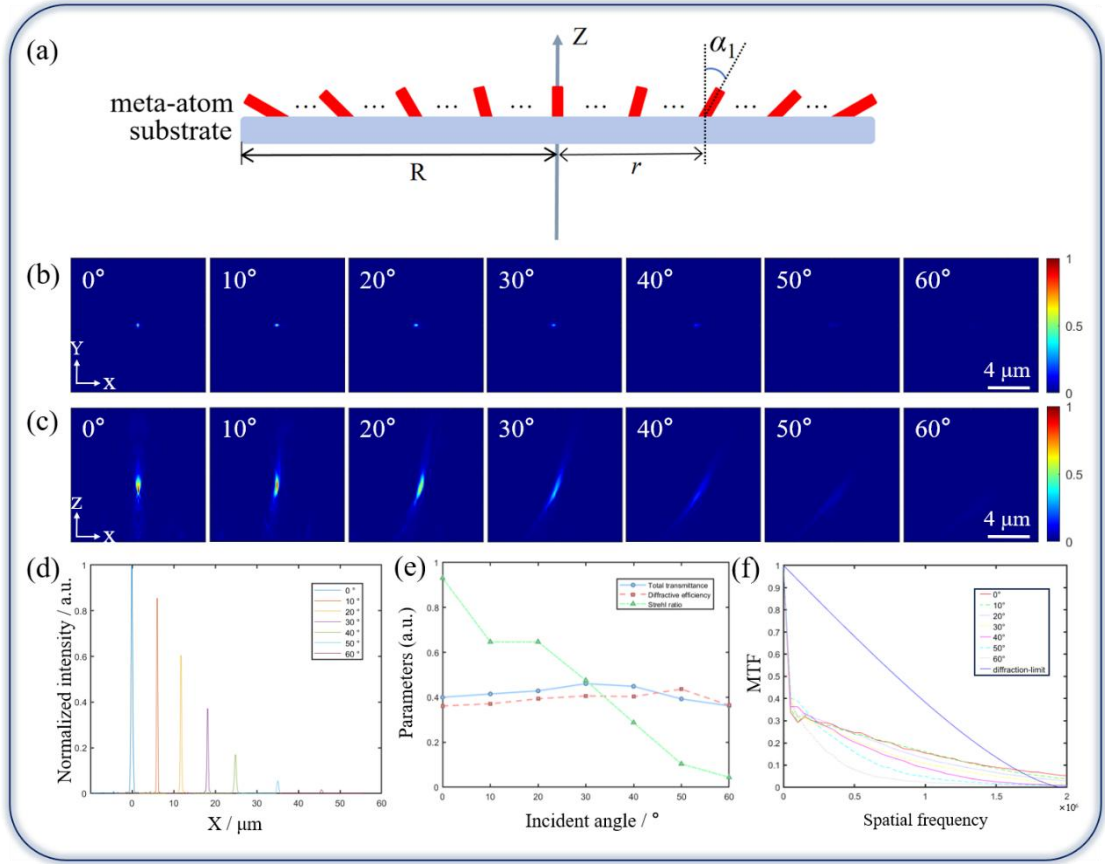

**Fig. 3.** WFOV metalens with PDTM. (a) Sectional schematic of meta-atoms arrangement on a substrate. Intensity distributions in (b) focal plane and (c) axial plane for incident angles of 0°, 10°, 20°, 30°, 40°, 50°, and 60°. (d) Normalized *x*-axis intensity profiles in the focal plane. (e) Total transmittance, diffraction efficiency, and Strehl ratio vs. incident angle. (f) The MTF of PDTM under different incident angles.

**Table 1** Parameter comparison between PDTM and several typical metalens designs.

| Design type | FOV (deg.) | Efficiency (%) | Effective NA | Wavelength (μm) |
|---|---|---|---|---|
| Singlet with an aperture stop[34] | 170 | 32 - 45 | 0.24 | 5.2 |
| Quadratic phase singlet[35] | 170 | 3.5 | 0.27 | 0.532 |
| Metalens doublet[36] | 50 | 50 | 0.44 | 0.532 |
| Metalens doublet[28] | 60 | 70 | 0.49 | 0.85 |
| **PDTM** | **120** | **38-43** | **0.625** | **0.633** |

## 3.2 Aperture-Dependent Tilted Meta-atom (ADTM)

While the PDTM prioritizes throughput, meta-atoms misaligned with the incident wavefront will introduce aberrations and limit the imaging precision correspondingly. To overcome this limitation, ADTM integrates a spatial filtering aperture with an optimized tilt-angle configuration, as shown in Fig. 1(a). The different colors are used to distinguish varying angles of light incidence. The aperture directs light to specific meta-atoms based on incident angle: orthogonally incident beams illuminate central meta-atoms, while large-angle rays target peripheral ones, leveraging rectilinear propagation. Fig. 4(a) presents a ray-tracing diagram illustrating this under normal and oblique incidence.

The tilt angle $α_2$ of meta-atoms is defined by the following geometric relationship:

$$α_2 = arctan\left(\frac{\sqrt{x^2+y^2}}{L}\right), \tag{4}$$

where $L$ is the aperture-to-metalens distance, and $x$ and $y$ are meta-atom coordinates. This creates a radial gradient of tilt angles, with peripheral meta-atoms having progressively larger $α_2$ values to correct wide-angle aberrations, as shown in Fig. 4(b). The diameter of designed ADTM is 51.6 μm with a focal length of 20 μm. So, this metalens possess a numerical aperture (NA) of 0.79. An aperture with radius of 6 μm is placed at the distance L = 17 μm from the metalens. Theoretically, the aperture placed in front of the lens will limit the maximum incident angle of light and reduce the effective NA correspondingly. In this work, the distance L between the aperture and the metalens is selected as an anchor point for the controlling of the tilt angles of the meta-atoms. Therefore, its impact on the effective NA can be neglected to achieve optimal focusing performance of the ADTM. To effectively filter out the incident light that would introduce additional aberrations while synchronously achieve a larger effective NA, the radius of aperture is set to the radius of the circular region occupied by meta-atoms whose tilt angles deviate by less than 10° from the direction of normally incident light. Based on this strategy, the controllable reduction in effective NA constitutes a beneficial influence for the metalens performance.

Similar to PDTM, we simulated ADTM's optical field distribution under X-polarized plane wave illumination at 633 nm for incident angles from 0° to 60°. Fig. 4(c-d) depict the intensity distributions in the focal and axial planes, respectively, demonstrating effective focusing with minimal stray light and sidelobes, with negligible background noise, ensuring high-contrast imaging. Additionally, the normalized intensity distribution for different incident angles of the focal plane along the X-axis are compared and shown in Fig. 4(e). The normalized light intensity remains 0.55 (relative to the 0° peak) at the incident angle of 30°. Additionally, the design achieves a substantial DOF averaging 13.4 μm, attributed to the phase gradient introduced by the tilted meta-atoms. This gradient not only mitigates minor spherical aberrations but also directly supports imaging over an extended focal range, enhancing robustness for wide-angle applications. Similarly, four critical metrics—total transmittance, diffraction efficiency, Strehl ratio, and MTF—are quantified to comprehensively evaluate the performance of ADTM, as shown in Fig. 4(f-g). Clearly, total transmittance decreases monotonically from 60% at 0° to 38% at 30°, reflecting reduced light throughput at larger angles due to weaker dipole response strength under high momentum mismatch conditions. However, the diffraction efficiency remains robust, averaging 0.8 from 0° to 50°, as the gradient-tilt architecture employs adiabatic phase matching to maintain efficient light coupling into the desired diffraction order. The Strehl ratio, a measure of focusing quality, declines from 0.92 at 0° to 0.41 at 60°, yet remains above 0.4 across the FOV, indicating effective aberration control compared to the PDTM's 0.48 at 30°. For spatial resolution, the MTF, plotted in Fig. 4(g) as a function of spatial frequency and incident angle, confirms diffraction-limited focusing from 0° to 30° and sub-diffraction performance at 40°–50°. This ensures consistent high-resolution imaging across a broad angular range, surpassing conventional metalenses that struggle with off-axis aberrations. To this end, we performed numerical simulations imaging the USAF resolution target and a complex pattern using this metalens, as shown in Fig. 4(h-i). The various fan-shaped regions in the figure represent imaging areas at different field angles, with the corresponding field angles labeled below. The smallest resolvable detail is approximately 1 μm. A complete and clear image can be formed across the entire FOV, with visibility gradually diminishing beyond the 60° field angle, though details remain distinctly resolved. These imaging results directly demonstrate ADTM's excellent WFOV, high-precision imaging capability.

  Overall, the ADTM's aperture-assisted, tilted meta-atom structure achieves a 120° FOV, delivering enhanced focusing precision and robust imaging capabilities for applications requiring wide-field coverage and tolerance to off-axis incidence.

Compared with conventional WFOV metalenses enabled by quadratic phase or multilayer structures, our design demonstrates superior coupling efficiency for large-angle incidence. Moreover, the radial gradient tilt configuration enables continuous WFOV operation through a single-layer metalens, eliminating the need for additional stitching or post-processing steps.

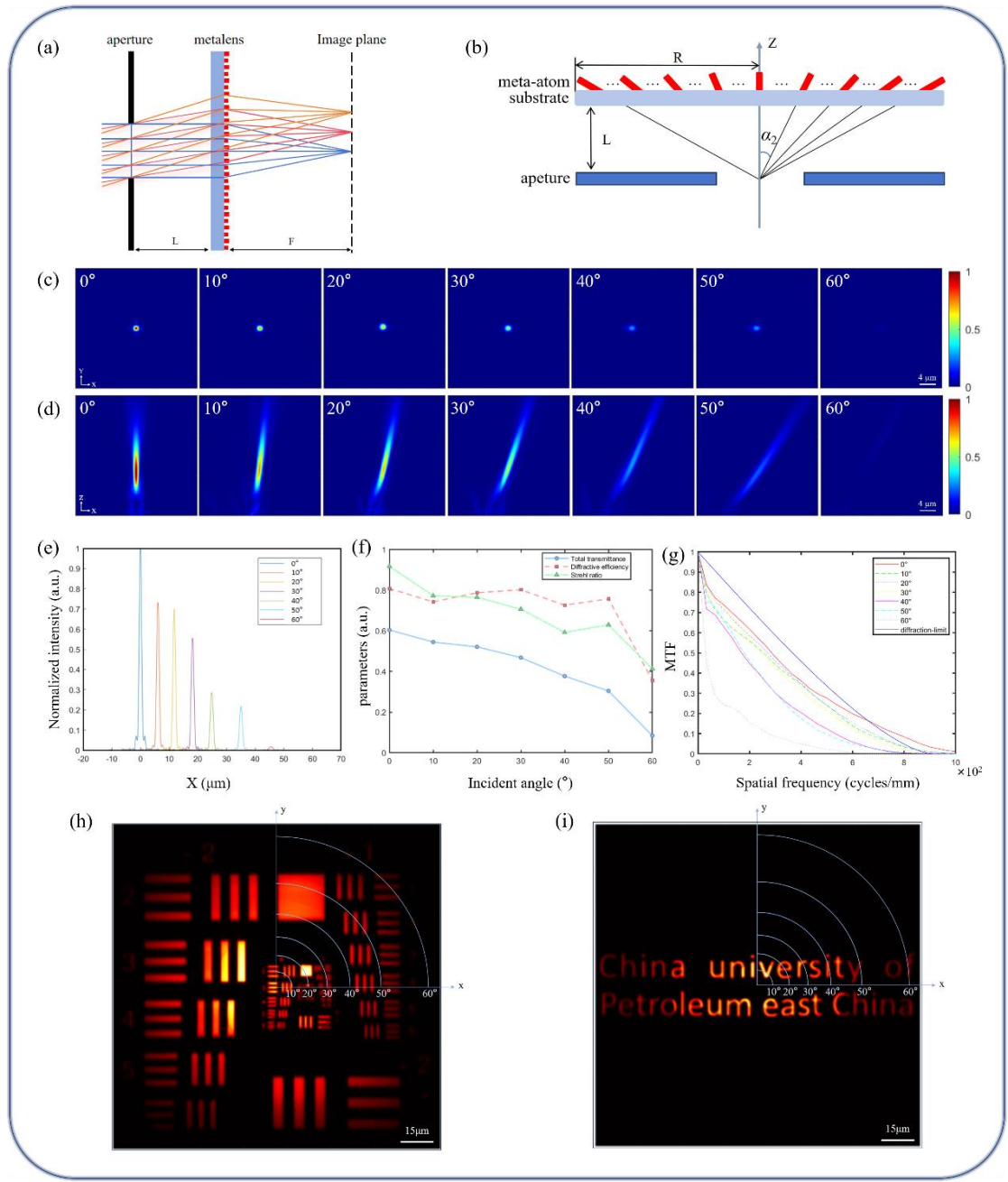

**Fig. 4.** WFOV metalens with ADTM. (a) Ray-tracing diagram of ADTM. (b) Sectional view of the ADTM. Intensity distributions in (c) the focal plane and (d) the axial plane of incident angles of 0°, 10°, 20°, 30°, 40°, 50°, and 60°. (e) Normalized *x*-axis intensity in the focal plane. (DOF). (f) Total transmittance, diffraction efficiency, and Strehl ratio vs. incident angle. (g) MTF vs. spatial frequency and incident angle. The simulation imaging of (h) the

USAF resolution target and (i) a complex pattern using ADTM.

Additionally, the proposed WFOV metalens in this work represents a proof-of-concept case and is far from the limit of this method. Larger FOVs are potentially achievable through tuning the tilt angle configurations of meta-atoms and metalens dimension. To demonstrate this, within the proposed design framework, we reconfigured the tilt angle arrangement of meta-atoms and the size of the metalens for the large FOV, and the focusing fields are shown in Fig. 6. Specifically, from the focal field of the ADTM shown in Fig. 6(a), it can be observed that although minimal distortion is still maintained at large incident, the efficiency drops inevitably. By contrast, the adjusted metalens shown in Fig. 6(b) maintains a better focusing efficiency from 50° to 70°. This fully indicates that our proposed WFOV metalens design framework, based on radial gradient tilting, exhibits FOV scalability.

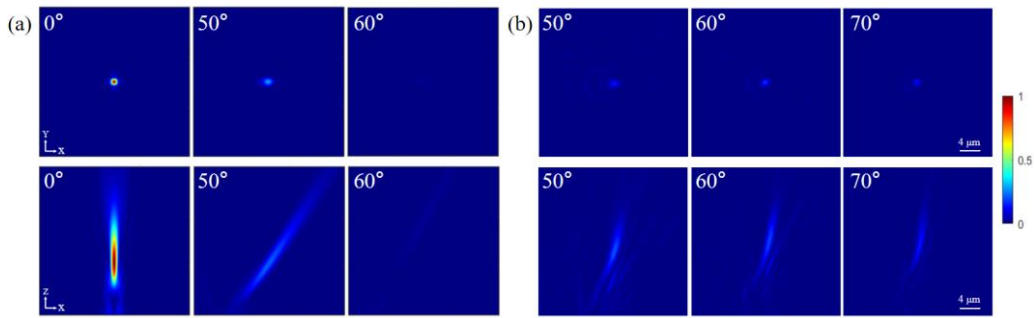

**Fig. 5.** The focusing field of the (a) ADTM and (b) the extended-FOV metalens.

## 4. Conclusion

This study introduces an innovative approach to wide-field-of-view (WFOV) metalens design, leveraging radially gradient-tilted meta-atom arrays and a spatial filtering aperture to achieve high-performance wide-angle imaging. We proposed and evaluated two complementary metalens designs: the Position-Dependent Tilted-Meta-Atom Metalens (PDTM) and the Aperture-Dependent Tilted-Meta-Atom Metalens (ADTM), analyzing their focal behavior, focusing efficiency, and imaging metrics through Finite-Difference Time-Domain simulations. The PDTM addresses the challenge of energy coupling efficiency degradation at large incident angles, maintaining robust transmittance (average 0.42) and diffraction efficiency (average 0.39) across a 120° FOV. The ADTM, with its aperture-assisted tilt configuration, achieves an average diffraction efficiency of 0.8 (0°–50°) and a Strehl ratio ranging from 0.92 at 0° to 0.41 at 60° (average ~0.67), while its extended depth of focus (13.4 µm) effectively mitigates spherical aberrations. Together, these designs enable diffraction-limited focusing through a single metasurface, delivering an uninterrupted 120° FOV that surpasses the limitations of conventional metalenses.

The PDTM and ADTM collectively resolve the trade-off between optical throughput and imaging precision, offering versatile solutions for diverse applications. Additionally, by adjusting the metalens diameter and reconfiguring the tilt angles of the meta-atoms, the FOV can be further extended, providing a scalable framework for future designs. These advancements pave the way for compact, high-resolution optical systems in fields such as wearable VR/XR displays, ultra-high-definition holography, and smart city infrastructure for traffic and environmental analytics. Our work establishes a new degree of freedom in metalens design, demonstrating significant potential for next-generation wide-angle imaging technologies.

**Acknowledgments**

This work was supported by National Key Research and Development Program of China (2023YFF1205704) and Western Youth Scholars Project of Chinese Academy of Sciences (No. XAB2022YN13). X.L. acknowledges the financial support from the Research Council of Finland (#354199 and #346529). The data presented in this study were generated independently and were not funded by the RCF.

**Conflict of interests**

The authors declare that they have no known competing financial interests or personal relationships that could have appeared to influence the work reported in this paper.